\documentclass[9pt,twocolumn,twoside]{pnas-new}
\usepackage{bm}

\templatetype{pnasmathematics} % Choose template 
\title{Atom2Vec: learning atoms for materials discovery}

\author[a]{Quan Zhou}
\author[a]{Peizhe Tang} 
\author[a]{Shenxiu Liu}
\author[b]{Jinbo Pan}
\author[b]{Qimin Yan}
\author[a,c,1]{Shou-Cheng Zhang}

\affil[a]{Department of Physics, McCullough Building, Stanford University, Stanford, California 94305-4045, USA}
\affil[b]{Department of Physics, Temple University, Philadelphia, Pennsylvania 19122, USA}
\affil[c]{Stanford Institute for Materials and Energy Sciences, SLAC National Accelerator Laboratory, Menlo Park, California 94025, USA}

\leadauthor{Zhou} 

\significancestatement{Motivated by the recent achievements of artificial intelligence (AI) in linguistics, we design AI to learn properties of atoms from materials data on its own. Our work realizes knowledge representation of atoms via computers for the first time, and could serve as a foundational step towards the materials discovery and design fully based on machine learning.}

\authorcontributions{Q.Z., P.T., S.L., Q.Y. and S.-C.Z. conceived and designed the project. Q.Z. designed and performed the model-free method of atom learning. Q.Z., P.T. and S.L. designed and performed the model-based method, proposed and performed supervised learning tasks for atom vector evaluations. J.P. and Q.Y. provided all data support in this work. Q.Z. and P.T analyzed all results and drafted the manuscript. All authors commented on the manuscript.}
\authordeclaration{The authors declare that they have no competing financial interests.}
\correspondingauthor{\textsuperscript{1}To whom correspondence should be addressed. E-mail: sczhang@stanford.edu}


\begin{abstract}
Exciting advances have been made in artificial intelligence (AI) during the past decades. Among them, applications of machine learning (ML) and deep learning techniques brought human-competitive performances in various tasks of fields, including image recognition, speech recognition and natural language understanding. Even in Go, the ancient game of profound complexity, the AI player already beat human world champions convincingly with and without learning from human. In this work, we show that our unsupervised machines (Atom2Vec) can learn the basic properties of atoms by themselves from the extensive database of known compounds and materials. These learned properties are represented in terms of high dimensional vectors, and clustering of atoms in vector space classifies them into meaningful groups in consistent with human knowledge. We use the atom vectors as basic input units for neural networks and other ML models designed and trained to predict materials properties, which demonstrate significant accuracy.
\end{abstract}

\dates{This manuscript was compiled on \today}
\doi{\url{www.pnas.org/cgi/doi/10.1073/pnas.XXXXXXXXXX}}

\begin{document}

% Optional adjustment to line up main text (after abstract) of first page with line numbers, when using both lineno and twocolumn options.
% You should only change this length when you've finalised the article contents.
\verticaladjustment{-2pt}

\maketitle
\thispagestyle{firststyle}
\ifthenelse{\boolean{shortarticle}}{\ifthenelse{\boolean{singlecolumn}}{\abscontentformatted}{\abscontent}}{}

% If your first paragraph (i.e. with the \dropcap) contains a list environment (quote, quotation, theorem, definition, enumerate, itemize...), the line after the list may have some extra indentation. If this is the case, add \parshape=0 to the end of the list environment.
\dropcap{T}he past twenty years witnessed the accumulation of unprecedentedly massive amount of data in materials science via both experimental explorations and numerical simulations\cite{Ceder2009,Norskov2009,Jain2013,Curtarolo2013,Choudhary2016}. The huge datasets not only enable but also call for data-based statistical approaches. As a result, a new paradigm emerges which aims to harness AI and ML techniques \cite{deepimage,deepspeech,deeplanguage,Silver2016,Silver2017} to assist materials research and discovery. Several initial attempts have been made along this path \cite{Muller2012,Wolverton2014,Rafael2016,Faber2016,Xue2016,Raccuglia2016}. Most of them learned maps from materials information (input) to materials properties (output) based on known materials samples. The input or feature of materials involves descriptors of constituents: certain physical or chemical attributes of atoms are taken, depending on the materials property under prediction\cite{Muller2012,Faber2015,Faber2016}. Despite the success so far, these works heavily rely on researchers' wise selection of relevant descriptors, thus the degree of intelligence is still very limited from a theoretical perspective. And practically, extra computations are usually unavoidable for machines to interpret such atom descriptors which are in the form of abstract human knowledge.

In order to create a higher level of AI and to overcome the practical limitation, we propose Atom2Vec in this letter, which let machines learn their own knowledge about atoms from data. Atom2Vec considers only existence of compounds in materials database, without reference to any specific property of materials. As far as we know, this is the first time that massive dataset is leveraged for learning feature in materials science in an unsupervised manner\cite{Hastie01,Bishop2006,LeCun2015,Goodfellow2016,Mikolov2013,pennington2014}. Because of the absence of materials property labels, Atom2Vec is naturally prevented from being biased to one certain aspect. As a result, the learned knowledge can yield complete and universal descriptions of atoms in principle, as long as the dataset is sufficiently large and representative. Atom2Vec follows the core idea that properties of an atom can be inferred from the environments it lives in, which is similar to the distributional hypothesis in linguistics\cite{Zelling1954}. In a compound, each atom can be selected as a target type, while the environment refers to all remaining atoms together with their positions relative to the target atom. Intuitively, similar atoms tend to appear in similar environments, which allows our Atom2Vec to extract knowledge from the associations between atoms and environments, and then represent it in a vector form as discussed in the following.

\section*{Atom2Vec Workflow}

\begin{figure*}[t]
\centering
\includegraphics[width=11.4cm]{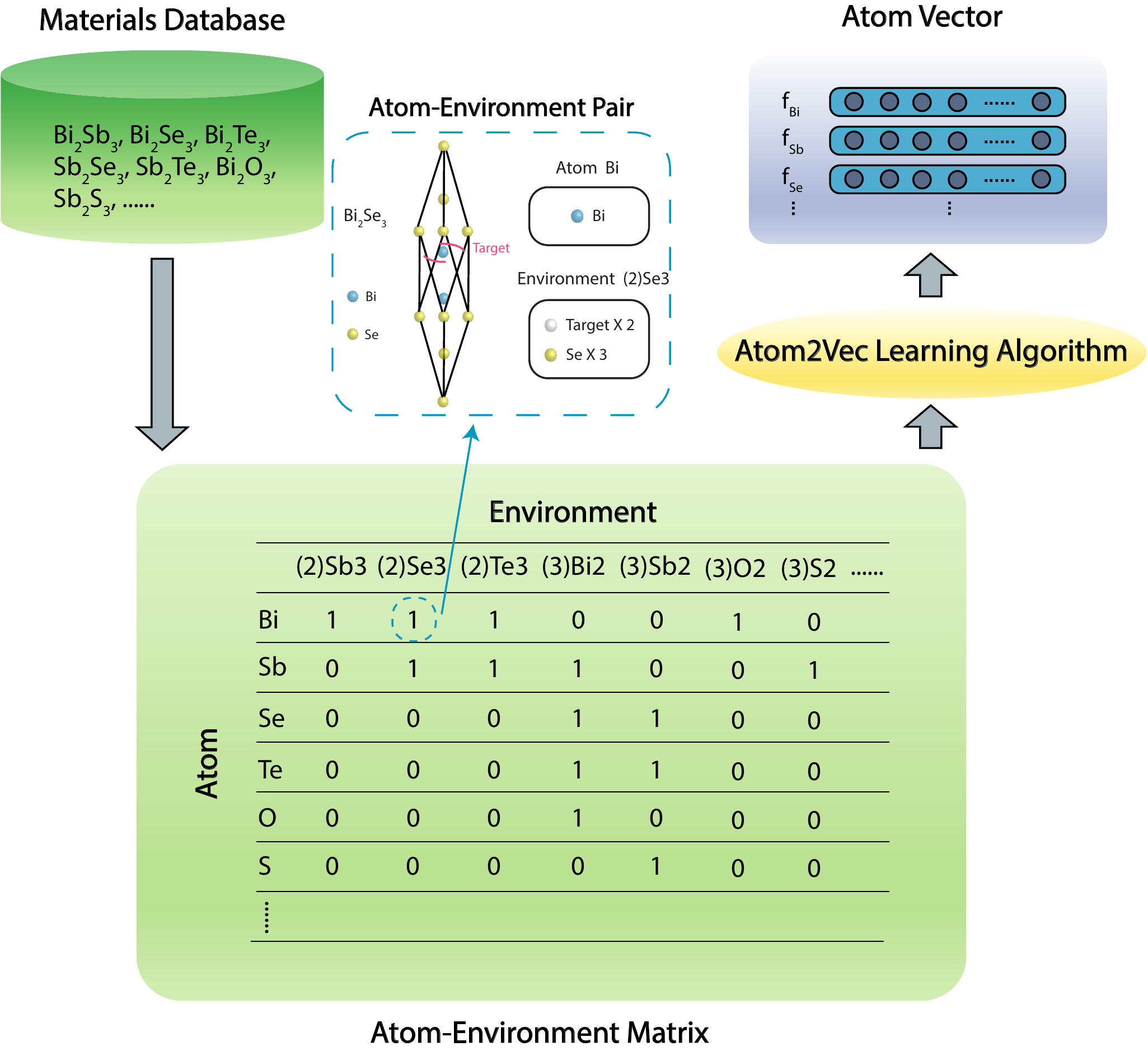}
\caption{\textbf{Atom2Vec workflow to learn atom from materials database.} Atom-environment pairs are generated for every compound in materials database, based on which atom-environment matrix is constructed. A small dataset of 7 compounds is used here as an example. Entries of atom-environment matrix denote the numbers of atom-environment pairs. Inset shows the unit cell of compound $Bi_2Se_3$, and the pair corresponding to the entry of target atom $Bi$ and environment $``(2)Se_3"$. Only compositional information is considered, while structural information is ignored. Atom2Vec learning algorithms extract knowledge of atoms from atom-environment matrix, and encode learned properties in atom vectors.}\label{fig:workflow}
\end{figure*}

We begin to illustrate the full workflow of Atom2Vec as shown in Figure \ref{fig:workflow}. In order to capture relations between atoms and environments, atom-environment pairs are generated for each compound in materials dataset as the first step. Before pair generation, a more explicit definition of environment is needed, whereas atoms are represented by chemical symbols conveniently. Although a complete environment should involve both chemical composition and crystal structure as mentioned before, we only take into account the former here as a proof of concept, and leave the latter for future study. Under this simplification, environment covers two aspects: the count of target atom in the compound, and the counts of different atoms in the remaining. As an example, let us consider the compound $Bi_2Se_3$ from the mini dataset of only seven samples given in Figure \ref{fig:workflow}. Two atom-environment pairs are generated from $Bi_2Se_3$: for atom $Bi$, the environment is represented as $``(2)Se3"$; for atom $Se$, the environment is represented as $``(3)Bi2"$. Specifically, for the first pair, $``(2)"$ in the environment $``(2)Se3"$ means there are two target atoms ($Bi$ here for the compound), while $``Se3"$ indicates that three $Se$ atoms exist in the environment. Following the notation, we collect all atom-environment pairs from the dataset, and then record them in an atom-environment matrix $\bm{X}$, where its entry $X_{ij}$ gives the count of pairs with the $i$th atom and the $j$th environment. Such a matrix for the mini dataset is also given in Figure \ref{fig:workflow} for illustration purpose. Clearly, each row vector gives counts with different environments for one atom, and each column vector yields counts with different atoms for one environment. According to the previously mentioned intuition, two atoms behave similarly if their corresponding row vectors are close to each other in the vector space.

Although revealing similarity to some extent, descriptions of atoms in terms of row vectors of atom-environment matrix are still very primitive and inefficient, since the vectors can be extremely sparse as every atom is usually related to only a small portion of all environments. To enable machines to learn knowledge about atoms beyond the raw data statistics, we have to design algorithms that learn both atoms and environments simultaneously. The idea originates from the observation that knowing environments (atoms) could help learn atoms (environments), where underlying high-level concepts can be distilled in the collaborative process. For example, in the mini dataset in Figure \ref{fig:workflow}, one can not directly conclude that $S$ and $O$ share attributes since there is no environment shared by them ($S$ only has environment $``(3)Bi2"$ while $O$ only has environment $``(3)Sb2"$ in the dataset). However, by comparing atoms associated to $``(3)Bi2"$ and $``(3)Sb2"$, one can find out that the two environments are similar to each other, which in turn indicates that $S$ and $O$ are actually similar. Herein we have two types of learning algorithms to realize the high-level concept or knowledge extraction from atom-environment matrix. The first type does not involve any modeling of how atom and environment interact with each other, thus is named model-free machine. In the model-free machines, singular value decomposition\cite{Hastie01} (SVD) are directly applied on re-weighted and normalized atom-environment matrix (see Materials and Methods for details), and row vectors in the subspace of the $d$ largest singular values encode the learned properties of atoms. The other type, model-based machine, assumes probability model about association between atom and environment, where randomly initialized $d$-dimensional vectors for atoms are optimized such that the likelihood of existing dataset is maximized (see Materials and Methods for details). We proceed to analyze learned atom vectors in the below. As our model-based machines are found to yield inferior vectors compared to model-free ones, probably due to the simplified modelings (see details in Materials and Methods), we focus on results from model-free learning in this work, whereas atom vectors from model-based machines are involved only for comparison when necessary (see details in Supplementary information).

\subsection*{Atom Vectors}

\begin{figure*}[t]
\centering
\includegraphics[width=17.8cm]{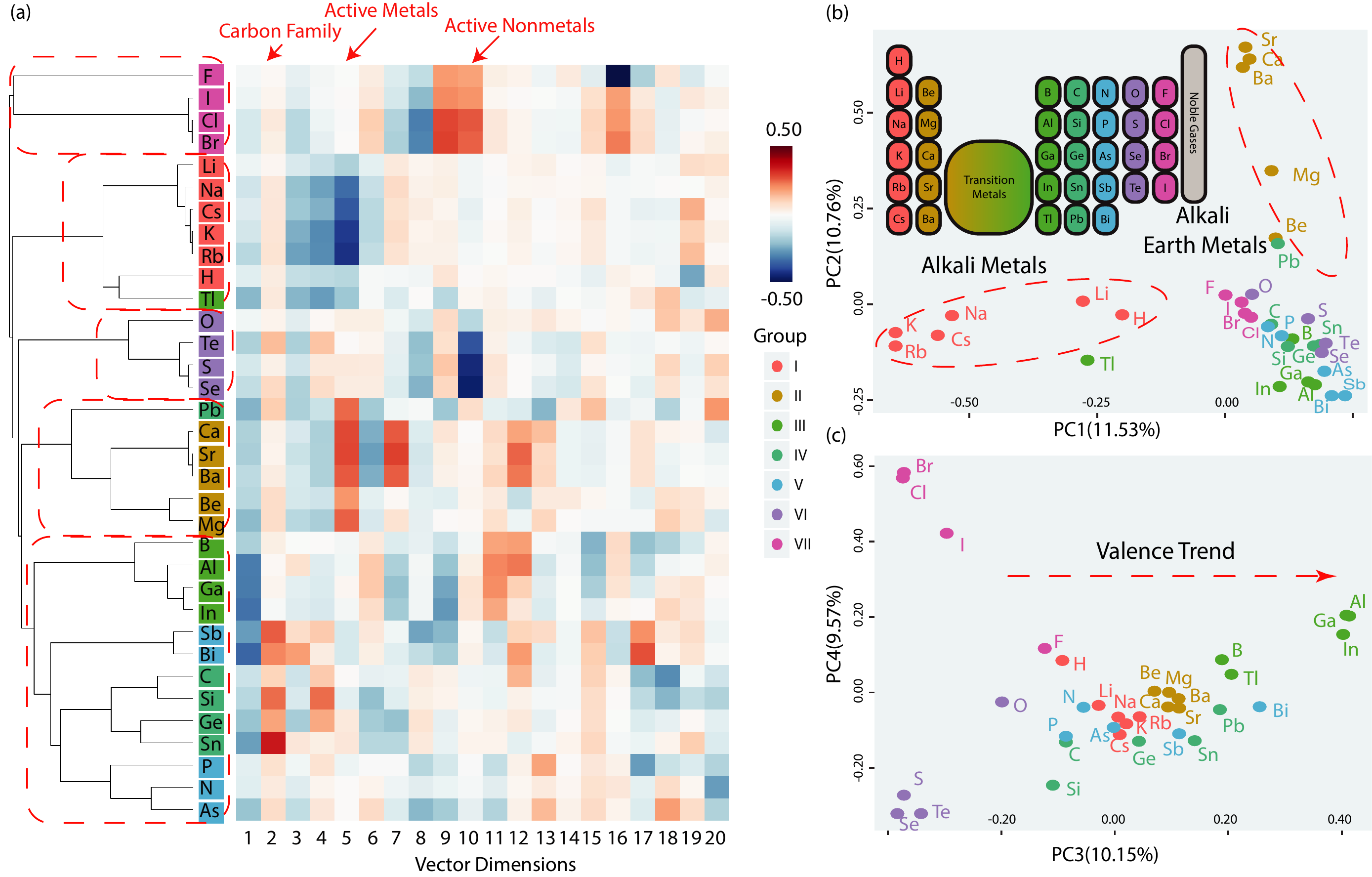}
\caption{ \textbf{Atom vectors of main-group elements learned by model-free approach.} (a) Illustration of atom vectors of 34 main-group elements in vector space of dimension $d=20$, and their hierarchical clustering based on distance metric $dist(\bm{f}_1, \bm{f}_2) = 1 - \bm{f}_1\cdot \bm{f}_2$ (see Materials and Methods). Row and column denote atom type and dimension index respectively; color in each cell stands for value of the vector on that dimension. Background colors of atom symbols label their columns in the periodic table. Dashed red boxes circle out major atom clusters from hierarchical clustering. Red arrows point to dimensions distinguishing different types of atoms. (b) Projection of the atom vectors of 34 main-group elements onto the plane spanned by the first and second principal axes (PC1 and PC2).  Percentage in the bracket gives the proportion of variance on that principal axis direction. Inset shows the periodic table of elements for reference. Dashed red circles show two distinctive clusters corresponding to two types of active metals. (c) Projection of atom vectors of 34 main-group elements onto the plane spanned by the third and fourth principal axes (PC3 and PC4). Percentage in the bracket gives the proportion of variance on that principal axis. Dashed red arrow indicates a valence trend almost parallel to PC3.}\label{fig:maingroup}
\end{figure*}

We firstly examine learned atom vectors for main-group elements, and show that they indeed capture atoms' properties. The atom vectors learned by our model-free machine are shown in Figure \ref{fig:maingroup}(a), together with the result of hierarchical clustering algorithm\cite{Hastie01} based on cosine distance metric in the vector space (see Materials and Methods for details). Cluster analysis is used to identify similar groups in data\cite{Hastie01}, and here we find that based on our atom vectors it manages to classify main-group elements into groups exactly the same as those in the periodic table of chemical elements\cite{Greenwood97}. Active metals including alkali metals (Group I) and alkali earth metals (Group II), and active nonmetals including chalcogens (Group VI) and halogens (Group VII) all reside in different regions in the high-dimensional vector space. Elements in the middle of the periodic table (Group III, IV and V) are clustered together into a larger group, indicating their similar properties. We also find in the clustering result that elements in high periods of the periodic table tend to be more metallic, for example, $Tl$ from Group III is close to alkali metals, and $Pb$ from Group IV is near alkali earth metals, both of which agree with chemical knowledge about these atoms. Moreover, there exist clear patterns in the heatmap layout of atom vectors in Figure \ref{fig:maingroup} (a), indicating that different dimensions of the vector space stand for different attributes of atoms. For instance, the 2nd dimension picks out carbon family, while the 5th dimension and the 10th dimension select out active metals and active nonmetals respectively. To better understand atom vectors in the high dimensional vector space, we project them into several leading principal components\cite{Hastie01} (PCs) and observe their distribution. As shown in Figure \ref{fig:maingroup} (b) and (c), PC1 and PC2 here convincingly separate alkali metals and alkali earth metals from others. Notably, PC3 bears a moderate resemblance to valence trend among these main-group elements.

\begin{figure*}[t]
\centering
\includegraphics[width=17.8cm]{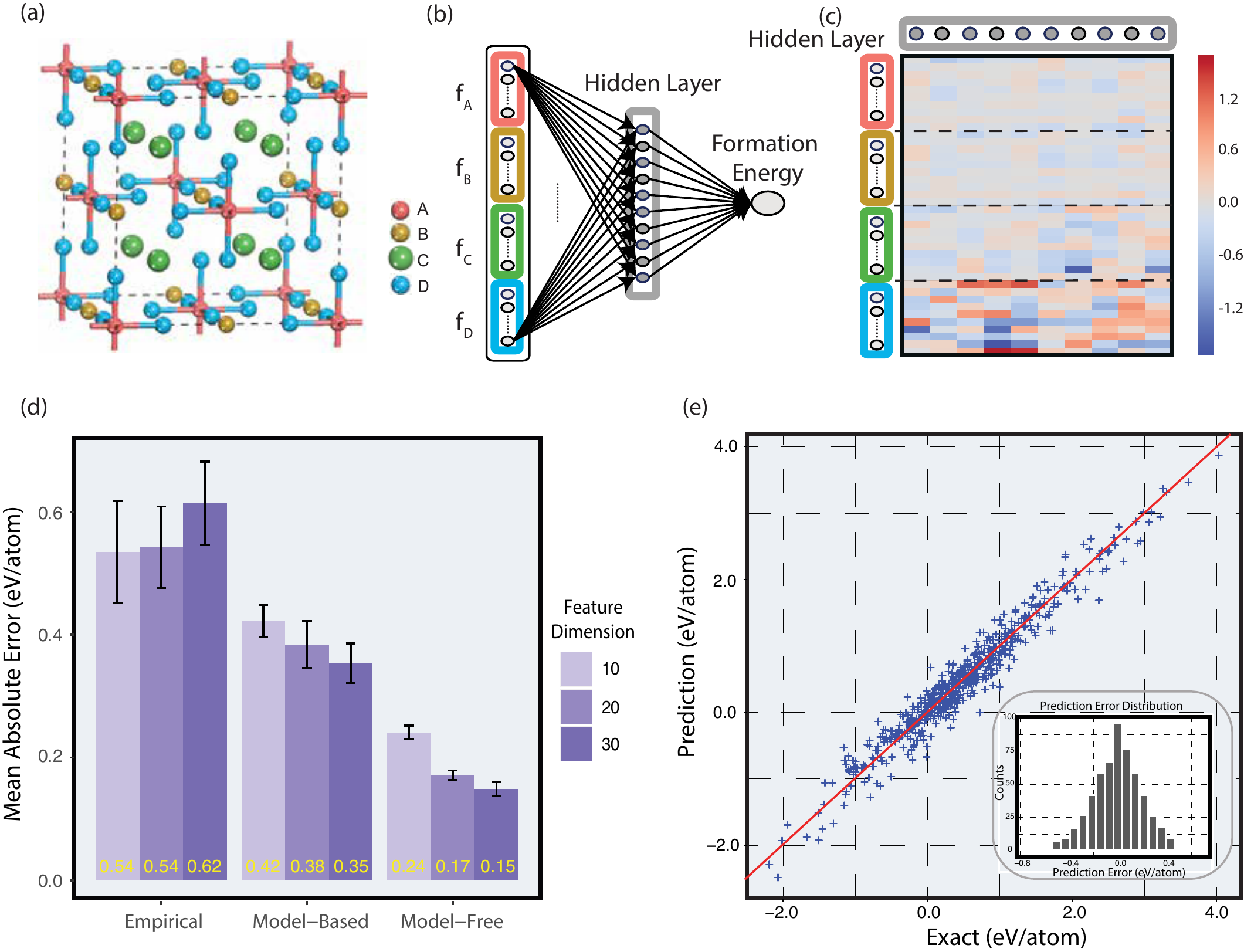}
\caption{ \textbf{Evaluation of atom vectors of main-group elements on elpasolites $\mathrm{ABC}_2\mathrm{D}_6$ formation energy prediction.} (a) Crystal structure of elpasolites $\mathrm{ABC}_2\mathrm{D}_6$. (b) Architecture of the one-hidden-layer neural network for formation energy prediction. Colored boxes represent atom vectors of atom $A$, $B$, $C$ and $D$ respectively, grey box in the hidden layer stands for representation of the elpasolites compound. (c) Trained weights on connections between the input layer and the hidden layer in the neural network for formation energy prediction using model-free atom vectors of dimension $d=10$. (d) Mean absolute test errors of formation energy prediction using different sets of atom features. Empirical features refer to the position of an atom in the periodic table, padded with random noise in expanded dimensions if necessary. Model-based features are atom vectors learned from our model-based method using inverse square score function (see Materials and Methods). Model-free features are atom vectors learned from our model-free method. Error bars show the standard deviations of mean absolute prediction errors on five different random train/test/validation splits. (e) Comparison of exact formation energy and predicted formation energy using $d=20$ model-free atom vectors. Inset shows the distribution of prediction errors.}\label{fig:formation}
\end{figure*}

After confirming that our atom vectors learn atoms' properties, we then verify that they are more effective in use for ML materials discovery than previously widely used knowledge-based descriptors of atoms (referred as empirical features in the following). We compare our atom vectors with those empirical features in a supervised learning task of materials prediction. The task we consider is to predict formation energies of elpasolite crystals $\mathrm{ABC}_{2}\mathrm{D}_{6}$, a type of quaternary minerals with excellent scintillation performance, which are thus very suitable for application in radiation detection\cite{Hawrami2016}. The dataset of this task contains nearly $10^4$ elpasolites with the first-principle calculated formation energies, and the task has been first introduced and modeled using kernel regression in a previous study\cite{Faber2016}. Here we employ a neural network with one hidden layer to model the formation energy (see Figure \ref{fig:formation} and Materials and Methods). The input layer is the concatenation of feature vectors of the four atoms ($A$, $B$, $C$ and $D$), which is then fully connected to a hidden layer equipped with non-linear activation. The hidden layer learns to construct a representation for each compound during the supervised learning process, based on which the output formation energy is predicted. For comparison, we fix the architecture of the neural network model, feed in different types of descriptors for atoms (empirical ones and atom vectors), and examine the prediction performances (see Materials and Methods). Like most previous studies, we use the positions of atoms in the periodic table as empirical features\cite{Faber2016}, but extend them to the same dimension as our atom vectors by random padding for fairness. Figure \ref{fig:formation} (d) shows the formation energy prediction errors based on empirical features, model-based learned atom vectors and model-free learned atom vectors. The latter two clearly yield higher prediction accuracies than the empirical one, which supports the superiority of the machine learned features over those from human knowledge. The influence of the dimension $d$ is also examined, and it turns out that larger $d$ leads to slight performance gain, as longer vectors could include more information. It is worth noting that, with our model-free atom vectors, the mean absolute error of formation energy prediction can be as low as 0.15eV/atom, almost within the error range of the first-principle calculation\cite{Matthias2015}. Figure \ref{fig:formation} (e) shows the predicted formation energies versus the exact values, and the error follows a normal distribution approximately. As a validation, we also check the neural weights between the input layer and the hidden layer in one model, which is visualized in Figure \ref{fig:formation} (c). The heavier weights for atom $\mathrm{D}$ indicates its dominant role in this problem, which agrees with previous results\cite{Faber2016}.

\begin{figure*}[t]
\centering
\includegraphics[width=17.4cm]{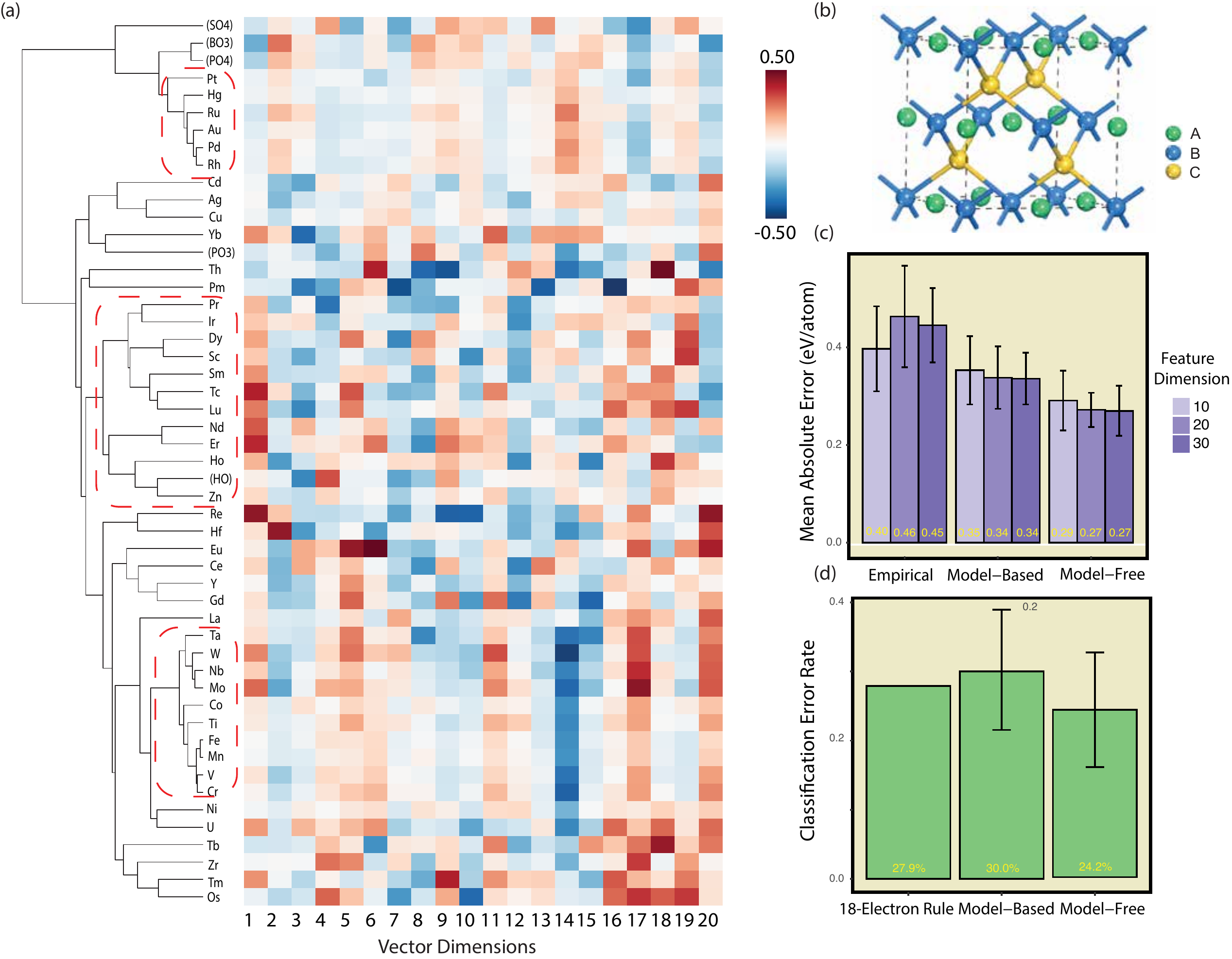}
\caption{\textbf{Atom vectors of functional groups and elements beyond main groups learned by model-free approach and the evaluation on tasks of half-heusler compounds.} (a) Illustration of atom vectors of non-main-group elements and functional groups in vector space of dimension $d=20$, and their hierarchical clustering based on distance metric $dist(\bm{f}_1, \bm{f}_2) = 1 - \bm{f}_1\cdot \bm{f}_2$ (see Methods). Row and column denote atom type and dimension indices respectively; color in each cell stands for value of the vector on that dimension. Dashed red boxes circle out major clusters in hierarchical clustering. (b) Crystal structure of half-heusler alloys $\mathrm{ABC}$. (c) Mean absolute test errors of formation energy prediction given by ridge regression using different sets of atom features. (d) Mean classification error rates of metal/insulator classifications given by the 18-electron rule, and logistic regression with model-based and model-free atom vectors of dimension $d=20$.}\label{fig:nonmaingroup}
\end{figure*}

Our learned atom vectors not only work well for main-group elements, but also provide reasonable descriptions for other atoms including transition metals as well as functional groups. These entities are in general more versatile in properties in contrast to main-group elements, therefore accurate description is difficulty even for humans. In this part, we focus on the learned vectors of these atoms, and show their advantage over empirical descriptors. These vectors along with the hierarchical clustering result is shown in Figure \ref{fig:nonmaingroup}(a). Roughly, there appear three major clusters in the vector space: the transition metal cluster with $Au$ as a representative ($Au$-like cluster), the cluster of lanthanoids and actinoids (lanthanoid cluster), and the transition metal cluster with $Fe$ as a representative ($Fe$-like cluster). The strong amplitudes on the 14th dimension indicate $Au$-like and $Fe$-like clusters of transition metals, while lanthanoid cluster show quite a lot of hot and cold spots on other dimensions. We also note that all atoms from $Fe$-like cluster share relatively strong amplitudes on the 5th dimension, which is very similar to those of Group II atoms mentioned previously. This accounts for the common $+2$ valence state adopted by these transition metals\cite{Greenwood97}. Moreover, the atom vectors are compared with empirical ones in prediction tasks on half-heusler alloys. This type of compounds have received lots of research attention, because their peculiar band structures enable tunable realization of topological insulators and semi-metals\cite{Chadov2010}. Our tasks here include both formation energy prediction and metal/insulator classification, which are two crucial steps for topological materials search. We have nearly 250 half-heusler compounds $\mathrm{ABC}$ under space group 216, along with their calculated formation energies and band gaps. Since the number of samples is limited, we use relatively simpler models here: ridge regression for prediction, and logistic regression for classification\cite{Hastie01}. As shown in Figure \ref{fig:nonmaingroup} (c), both model-free and model-based atom vectors outperform empirical ones in formation energy prediction, and the best set of vectors achieves mean absolute error 0.27eV/atom. In metal/insulator classification as shown in Figure \ref{fig:nonmaingroup} (d), we compare the logistic regression model using our atom vectors with the famous 18-electron rule\cite{Chadov2010}. Comparable accuracies can be achieved here, which again supports the effectiveness of atom vectors in the ML tasks.

\subsection*{Summary and Outlook}

In summary, we introduce unsupervised learning of atoms from database of known existing materials, and show the rediscovery of the periodic table by AI. The learned feature vectors not only well capture similarities and properties of atoms in a vector space, but also show their superior effectiveness over simple empirical descriptors when used in ML problems for materials science. While empirical descriptors are usually designed specifically for a task, our learned vectors from unsupervised learning should be general enough to be applied to many cases. We anticipate their effectiveness and broad applicability can greatly boost the data-driven approaches in today's materials science, especially for the recently proposed deep neural network methods\cite{Kearnes2016,Dahl2017,Faber2017,Schutt2017}, just as the huge success of word vectors in language modeling\cite{Landauer1997,Mikolov2013,pennington2014,Socher2013}. Several directions related to the feature learning methods here are worthy to explore in the future. For example, the element-environment matrix can be generalized to a higher order tensor, where the extra orders depict different parts of the composition. Such a tensor should contain finer information than the matrix, and how to extract features from this high order object is still an open question. Also, more reasonable environment descriptions are necessary for improvement in both model-free and model-based methods. Structural information has to be taken into account in order to accurately model how atoms are bound together to form either environment or compound, where the recent development on recursive and graph-based neural networks\cite{Socher2013,Dahl2017,Schutt2017} might help.

\matmethods{ \subsection*{Data preprocessing}
All inorganic compounds from Materials Project database\cite{Jain2013} are used for our unsupervised feature learning of atoms. Compounds including more than four types of elements (or symbols in the following, to include functional groups in more general sense) are screened out for simplicity; only binary, ternary and quaternary compounds are selected, and they comprise nearly $90\%$ of the entire dataset of inorganic compounds (about 60k) (see Supplementary Information). There exist compounds appearing more than once in the database, all of them are kept since duplications give higher confidence which is meaningful in learning. No physical or chemical properties of compounds in the database are used, and as mentioned in the main text, the structural aspect are further ignored for simplicity. In other words, we build feature vectors of atoms merely from chemical formulas of all existing compounds. Some symbols are rare in terms of their number of appearances in compounds; they contribute very limited information and could impair feature learning due to high variance. So we collect counts of all symbols in compounds, and consider only common ones whose counts are beyond the specified threshold ($1\%$ of the maximum count of a symbol). Several atom-environment pairs can be generated from one compound formula. Symbols of the same type are assumed to be the same in a compound, so "atom" here literally represents the type of the target atom, and "environment" includes the number of target atoms and atoms of all other elements in the compound. Therefore the number of generated atom-environment pairs is equivalent to the number of symbol types in the compound formula.

\subsection*{Model-free methods}
Our model-free methods are based on atom-environment matrix built from atom-environment pairs. As the first step, we scan all the pairs and count the number of pairs where the $i$th atom and the $j$th environment appear together. These counts give the entry of atom-environment matrix $X_{i,j}$, where $i$ ranges from $1$ to $N$ while $j$ is between $1$ to $M$. Typically, the number of atom types $N$ is about one hundred, and the number of environments $M$ takes the order of tens of thousands, depending on the dataset ($N=85$ and $M=54032$ in our case). The sum of counts over the column $\sum_{j}X_{i,j}$ gives the population of the $i$th atom, which can differs greatly among all symbols. In order to get rid of the influence of such imbalance, we focus on common symbols as mentioned above, and apply the normalization $\mathcal{X}_{i,j} = X_{i, j} / (\sum_{j} X_{i, j}^p)^{\frac{1}{p}}$ on row vectors, where $p$ is a integer hyper-parameter that tunes the relative importance of rare environments with respect to common ones. In this way, larger $p$ emphasizes more the role of common environments while smaller $p$ tends to put equal attention on every environment. We test different choices of $p$ and select $p=2$, as it provides a natural distance metric in the vector space: the inner product of two normalized vectors, equivalently the cosine of the angle spanned by the two directions, denotes the similarity of the two symbols because it corresponds to a matching degree on all environments. The distance metric of the pair of normalized vectors $\bm{u}_1$ and $\bm{u}_2$ is explicitly defined as $dist(\bm{u}_1, \bm{u}_2) = 1 - \bm{u}_1\cdot\bm{u}_2$.

The row vectors of the normalized matrix $\bm{\mathcal{X}} = [\bm{x}_1, \bm{x}_2, ..., \bm{x}_N]^{T}$ provide a primitive representation for atoms. Essentially, each vector $\bm{x}_i$ gives the distribution profile over all environments, and vectors of similar atoms are close from one another in the high dimensional space because they are likely to appear in same environments. In order to have more efficient and interpretable representation for atoms, we apply singular value decomposition (SVD) on the normalized matrix $\bm{\mathcal{X}}$, and project original row vectors to dimensions of leading singular values. Specifically, we factorize the $N\times M$ matrix as $\bm{\mathcal{X}} = \bm{U} \bm{D} \bm{V}^{T}$, where $\bm{U}$ is $N\times N$ orthogonal matrix, $\bm{V}$ is $M\times M$ orthogonal matrix, and $\bm{D}$ is $N\times M$ is diagonal matrix with diagonal elements corresponding to singular values.  We select $d$ largest singular values (see Supplementary Information) from $\bm{D}$, say $d\times d$ matrix $\tilde{\bm{D}}$, and the corresponding columns from $\bm{U}$, namely $N\times d$ matrix $\tilde{\bm{U}}$; the product of the two matrices yields a $N\times d$ matrix, whose row vectors yield better descriptions for atoms
\begin{gather}
\bm{F} = \tilde{\bm{U}} \tilde{\bm{D}} = [\bm{f}_1, \bm{f}_2, ..., \bm{f}_N]^{T}.
\end{gather}
In contrast to the primitive representations, these features vectors $\bm{f}_i$ are more compact and describe elements in an abstract way. As another hyper-parameter, The dimension $d$ of the feature vectors is selected by a threshold on singular values (See Supplementary Information).

Because SVD almost preserves the structure of inner product, the distance metric mentioned previously also applies to our atom vectors. Therefore we perform hierarchical clustering analysis of these atom vectors of dimension $d$ based on this metric. Hierarchical clustering is used to build a hierarchy of clusters of the atoms according to a distance measure or similarity measure. Here we take a bottom-up approach in this work. In details, every atom vector starts in its own cluster, and the distance value is initialized as zero. Then the distance value is gradually increased, and when the distance of two clusters is smaller than the value, they are merged together as a new cluster. We take an single linkage approach here, namely the minimal distance between vectors from two clusters is used as the distance between the clusters. Thus, as the distance value becomes larger, all atoms find out the clusters they belong to and they are merged into one big cluster eventually. The resulted dendrogram shows the merge process and the clustering of atoms with similar properties.

\subsection*{Model-based methods}
Our model-based methods rely on more assumptions about representations of environments, besides atom-environment pairs. We represent each environment in the same vector space of atom vectors, and the composition model maps the collection of all atoms in the environment to a feature vector for the environment. Suppose feature vectors of atoms are of dimension $d$, which are given by $N\times d$ matrix $\bm{F} = [\bm{f}_1, \bm{f}_2, ..., \bm{f}_N]^{T}$. Consider one atom-environment pair whose environment includes $k$ atoms (here $k$ is the number of atoms except the single target atom), the atom type indices of these atoms are $i_1, i_2, ..., i_k$. Based on the assumed composition model $C$, the environment is represented as 
\begin{gather}
\bm{f}^{env} = C(\bm{f}_{i_1}, \bm{f}_{i_2}, ..., \bm{f}_{i_k}).
\end{gather}
As an example, for the environment $(2)Se3$ from compound $Bi_2Se_3$, $k=4$, namely the environment is composed of one $Bi$ and three $Se$, and the environment is thus
\begin{gather}
\bm{f}^{env} = C(\bm{f}_{Bi}, \bm{f}_{Se}, \bm{f}_{Se}, \bm{f}_{Se}).
\end{gather}
There are many choices for $C$ to fully characterize compositions, for example, recursive neural networks and graph-based models\cite{Socher2013,Schutt2017}. In this work, we select the composition model $C$ to be a summation over all atoms $\bm{f}^{env} = \sum_k \bm{f}_{i_k}$. Though this choice is oversimplified, it seems to already capture some major rules in composition (see Supplementary Information). Thus it is used here as a proof of concept, and more flexible and general composition models are worthy to test in the future. 

Another component in our model-based methods is a score function that evaluates the existence likelihood of an atom-environment pair. The score function $S$ received an atom vector $\bm{f}_{i}$ and a environment vector $\bm{f}^{env}$ as built above, and then returns a non-negative value $S(\bm{f}_{i}, \bm{f}^{env})$ indicating the probability for the pair to exist. A larger score means that such a pair is more likely to appear. Several score functions are designed following the basic intuition that an atom and an environment should behave oppositely in order to be combined as a stable compound. These are the bilinear score $S(\bm{f}_{i}, \bm{f}^{env}) = exp(-\bm{f}_{i}\cdot \bm{f}^{env})$, the Gaussian-like score $S(\bm{f}_{i}, \bm{f}^{env}) = exp(-||\bm{f}_{i} + \bm{f}^{env}||^2)$, and the inverse square score $S(\bm{f}_{i}, \bm{f}^{env}) = \frac{1}{||\bm{f}_{i} + \bm{f}^{env}||^2}$. 

In practice, a normalized score is assigned to each atom type given an environment
\begin{gather}
s_i = \frac{S(\bm{f}_i, \bm{f}^{env})}{\sum_l S(\bm{f}_l, \bm{f}^{env})}.
\end{gather}
The desired atom vectors need to maximize the average normalized score over the full dataset of atom-environment pairs, or minimize the following loss function
\begin{gather}
\textit{loss} = \mathbb{E}_{dataset} [-\ln s(\bm{f}_{i_e}, C(\bm{f}_{i_1}, \bm{f}_{i_2}, ..., \bm{f}_{i_k}))],
\end{gather}
where $\bm{f}_{i}$ stands for the target atom in each pair, and $\bm{f}_{i_1}, \bm{f}_{i_2}, ..., \bm{f}_{i_k}$ are for atoms in the environment. So our model-based feature learning is now cast into a optimization problem. To solve it, we firstly randomly initialize all feature vectors of atoms, then mini-batch stochastic gradient descent method is applied to minimize the loss function (see Supplementary Information). We choose the batch size to be 200, the learning rate to be 0.001, and the number of training epochs to be around one hundred. Feature vectors of different dimensions $d$ are learned following the model-based methods.

\subsection*{Models in prediction tasks}
We briefly introduce the machine learning methods used in our feature evaluation tasks. Neural network model is used to predict the formation energy of elpasolites $\mathrm{ABC}_{2}\mathrm{D}_{6}$. Initially inspired by neural science, in neural networks, artificial neurons are arranged layer by layer, and adjacent layers are fully connected with a linear transformation. Non-linearity function (such as sigmoid unit and rectified grated linear unit Relu) are applied on neurons on intermediate layers. Prediction, either regression or classification, is made according to the output layer. The weights and bias between layers are optimized in order to minimize the loss function over the training dataset, and the loss function over a hold-out dataset is used as validation that prevents overfitting and guarantees model generalizability. We train a neural network with one hidden layer for formation energy prediction in this work. The input layer $\bm{l}$ is a concatenation of the feature vectors of the four atom types in the compound, and its dimension is $4d$ where $d$ is the dimension of feature vectors. The hidden layer $\bm{h}$ contains $10$ neurons, which produces a representation of the compound. Formation energy is given by a weighted linear combination of these hidden units. Explicitly, the model is written as 
\begin{gather}
\bm{h} = Relu(\bm{W}\cdot \bm{l} + \bm{b}), \\
E_{formation} = \bm{w}\cdot\bm{h} + e,
\end{gather}
where $\bm{W}$ and $\bm{b}$ are weights and bias between the input layer and the hidden layer, $\bm{w}$ and $e$ are weights and bias between the hidden layer and output layer. Rectified linear unit function $Relu(z) = \Theta(z)z$ gives nonlinear activation here, where $\Theta(z)$ is Heaviside step function. The mean square error of formation energy prediction is used as loss function. There are 5645 $\mathrm{ABC}_2\mathrm{D}_6$ compounds with known formation energies, we randomly hold out 10 percents as test set. 80 percents as training set, and the other 10 percents as validation set. The training is terminated once the validation loss does not decrease any more; the trained model is then evaluated on the hold-out test set. We train and evaluate the model on 10 different random train-test splits, and the average of mean absolute errors on test sets is reported for each set of features.

For the task for half-heusler compounds, ridge regression and logistic regression are adopted. Ridge regression is a variant of vanilla linear regression that adds a $L2$ regularization term\cite{Hastie01} to the original mean square error loss function. This term prevents overfitting the training data, thus improves generalizability of the model. Logistic regression generalizes linear regression into a method for binary category classification. Sigmoid function is applied on the output unit, and a log-likelihood loss function replaces the mean square error loss. We apply ridge regression and logistic regression to formation energy prediction and metal/insulator classification respectively. The same data splits as above are taken, we report the average mean absolute error and classification error rate for models using different set of features.
}
\showmatmethods{} % Display the Materials and Methods section

\acknow{Q.Z., P.T. and S.-C.Z. acknowledge the Department of Energy, Office of Basic Energy Sciences, Division of Materials Sciences and Engineering, under contract DE-AC02-76SF00515. J. P. and Q. Y. are supported by the Center for the Computational Design of Functional Layered Materials, an Energy Frontier Research Center funded by the U.S. Department of Energy, Office of Science, Basic Energy Sciences under Award No. DE-SC0012575. }

\showacknow{} % Display the acknowledgments section

% \pnasbreak splits and balances the columns before the references.
% Uncomment \pnasbreak to view the references in the PNAS-style
% If you see unexpected formatting errors, try commenting out \pnasbreak
% as it can run into problems with floats and footnotes on the final page.
%\pnasbreak

% Bibliography
\bibliography{pnas-sample}

\end{document}